\documentclass[aps,prd,superscriptaddress,showpacs]{revtex4}

\usepackage{graphicx}
\usepackage{times}
\usepackage{dcolumn}
\newcommand{\bq}{\mathbf{q}}

\newcommand{\fp}{\langle F' \rangle}
\newcommand{\fpp}{\langle F'' \rangle}
\newcommand{\cs}{\mathcal{S}}
\newcommand{\csb}{\bar{\mathcal{S}}}

\begin{document}

\title{Reconstructing baryon oscillations}

\author{Yookyung Noh}
\email{ynoh@astro.berkeley.edu}
\affiliation{Department of Astronomy, 601 Campbell Hall,
University of California Berkeley, CA 94720}

\author{Martin White}
\email{mwhite@berkeley.edu}
\affiliation{Departments of Physics and Astronomy, 601 Campbell Hall,
University of California Berkeley, CA 94720}

\author{Nikhil Padmanabhan}
\email{nikhil.padmanabhan@yale.edu}
\affiliation{Dept. Physics, Yale University, New Haven, CT 06511}

\date{\today}

\begin{abstract}
The baryon acoustic oscillation (BAO) method for constraining the expansion
history is adversely affected by non-linear structure formation, which washes
out the correlation function peak created at decoupling.  To increase the
constraining power of low $z$ BAO experiments, it has been proposed that
one use the observed distribution of galaxies to ``reconstruct'' the acoustic
peak.  Recently Padmanabhan, White \& Cohn provided an analytic formalism for
understanding how reconstruction works within the context of Lagrangian
perturbation theory.  We extend that formalism to include the case of biased
tracers of the mass and, because the quantitative validity of LPT is
questionable, we investigate reconstruction in N-body simulations.
We find that LPT does a good job of explaining the trends seen in simulations
for both the mass and for biased tracers and comment upon the implications
this has for reconstruction.
\end{abstract}


\maketitle
\twocolumngrid

\section{Introduction}

It has been known for many years that the coupling of photons and baryons
in the early universe results in significant features in the matter power
spectrum \cite{PeeYu70,SunZel70,DorZelSun78}.
Prior to recombination, photons and baryons are tightly coupled and are
well approximated by a fluid. Perturbations during this epoch do not grow,
but instead excite sound waves which get frozen at recombination and
manifest themselves as an  almost harmonic series of peaks in the power
spectrum, $P(k)$, or equivalently a narrow feature in the correlation
function, $\xi(r)$
(see \cite{EisHu98,MeiWhiPea99} for a detailed description of the physics in
 modern cosmologies and \cite{ESW07} for a comparison of Fourier and
 configuration space pictures).
These so-called ``baryon acoustic oscillations'' (BAO) can be used as a
standard ruler to measure the expansion rate of the Universe, making the
method an integral part of current and next-generation dark energy experiments.

While the early Universe physics is linear and well understood, the low
redshift observations are complicated by the non-linear evolution of 
matter, galaxy bias and redshift space distortions.
The non-linear evolution leads to a coupling of $k$-modes and damping of
the oscillations on small scales \cite{MeiWhiPea99} and a small shift in
their positions \cite{ESW07,CroSco08,Mat08a,Seo08,PadWhi09}.
The damping of the linear power spectrum (or equivalently the smoothing of
the correlation function) reduces the contrast of the feature and 
the precision with which the size of ruler may be measured.

In \cite{ESW07} it was pointed out that much of the modification to the
power spectrum comes from bulk flows and super-cluster formation.
Since these large-scale flows are reasonably well measured by the survey,
their effects can, in principle, be corrected.
In \cite{ESSS07} a method was introduced for removing the non-linear
degradation of the acoustic signature, sharpening the feature in configuration
space or restoring/correcting the higher $k$ oscillations in Fourier space;
this method has been tested on simulations by a number of groups
\cite{ESSS07,HSWSW07,WagMulSte08}.
However, this method is inherently non-linear and therefore difficult to
understand analytically.
A study of this problem for the matter density using Lagrangian perturbation
theory \cite{PadWhiCoh09} explained how the method ``reconstructed'' the
BAO feature, but also pointed out that it did not reconstruct the linear
density field. We extend these results here - (i) generalizing the analytic
theory to biased tracers, including explicit expressions for the reconstructed
power spectrum to second order in the linear power spectrum, and (ii) testing
the validity of the analytic expressions with a suite of N-body simulations.

We compare the analytic theory to a set of $1024^3$ particle simulations run
in periodic, cubical boxes of side length $2\,h^{-1}$Gpc with a {\sl TreePM\/}
code \cite{TreePM}.  The simulations were initialized at $z=100$ using second
order Lagrangian perturbation theory, and the phase space information for
all of the particles was dumped at $z=0$, $0.3$, $0.7$ and $1.0$.
Multiple realizations, with different initial density fields, were run for
each cosmology to reduce sampling effects (more details can be found in
\cite{CarWhiPad09,PadWhi09}).
In addition to the dark matter particle data, halo catalogs were produced for
each output using the friends-of-friends method \cite{DEFW} with a linking
length of $0.168$ times the mean inter-particle spacing.  We work with halos
above $10^{13}\,h^{-1}M_\odot$, i.e.~containing more than 20 particles.

We investigate one of the cosmologies considered in \cite{CarWhiPad09,PadWhi09}:
$\Lambda$CDM, with $\Omega_M=0.25$, $\Omega_B=0.04$, $h=0.72$, $n=0.97$ and
$\sigma_8=0.8$.  This is close to the current ``best fit'' cosmology and will
serve as a realistic model to explore.  Within this cosmology the acoustic peak
in the correlation function is at $\sim 110\,h^{-1}$Mpc, with an intrinsic
width set by the diffusion (Silk) damping scale of $\sim 10\,h^{-1}$Mpc.

\section{Reconstruction I: Matter}

We begin our investigation of reconstruction by considering the most
conceptually simple case: reconstruction of the acoustic peak in the
matter 2-point function. We start by reviewing the reconstruction algorithm
of \cite{ESSS07} and its interpretation within Lagrangian perturbation 
theory \cite{PadWhiCoh09}, and then compare its predictions with simulations.
The following section extends this analysis to biased tracers.

\subsection{Algorithm}

\begin{figure}
\begin{center}
\resizebox{3.5in}{!}{\includegraphics{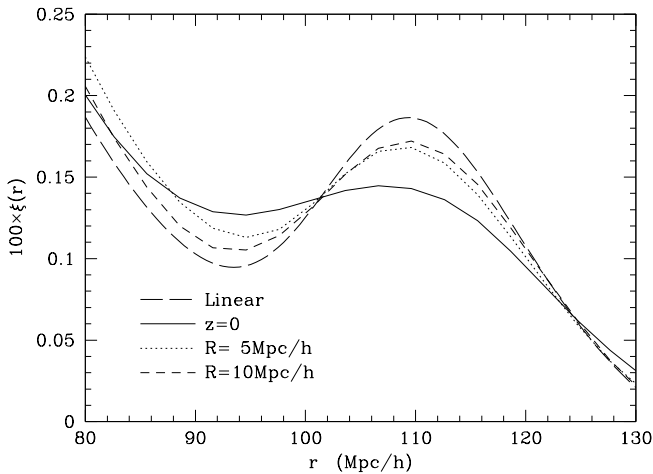}}
\end{center}
\vspace{-0.2in}
\caption{The mass correlation function for our $\Lambda$CDM model at $z=0$
before (solid) and after reconstruction using a smoothing of
$R=5\,h^{-1}$Mpc (dotted) and $R=10\,h^{-1}$Mpc (short dashed).  Non-linear
evolution has partially erased the peak in the initial conditions
(long-dashed) by $z=0$, but it is somewhat restored by reconstruction.}
\label{fig:example}
\end{figure}

The algorithm devised by \cite{ESSS07} is straightforward to apply to a
simulation and consists of the following steps:
\begin{itemize}
\item Smooth the density field to filter out high $k$ modes, which are
difficult to model.
\item Compute the negative Zel'dovich displacement, $\mathbf{s}$, from the
smoothed density field:
$\mathbf{s}(\mathbf{k})=-i(\mathbf{k}/k^2)\delta(\mathbf{k}) {\cal S}(k)$,
where ${\cal S}$ is the smoothing kernel (see below).
\item Shift the original particles by $\mathbf{s}$ and compute the ``displaced''
density field, $\delta_d$.
\item Shift an initially spatially uniform distribution of particles by
$\mathbf{s}$ to form the ``shifted'' density field, $\delta_s$.
\item The reconstructed density field is defined as
$\delta_r\equiv \delta_d-\delta_s$ with power spectrum
$P_r(k)\propto \langle \left| \delta_r^2\right|\rangle$.
\end{itemize}
Following \cite{ESSS07} we use a Gaussian smoothing of scale $R$,
specifically
\begin{equation}
  {\cal S}(k) =  e^{-(kR)^2/2}
  \quad .
\end{equation}
We take advantage of the periodicity of the simulations to perform all
of these steps using fast Fourier transforms.  The density fields are
constructed from the particle positions using a CIC assignment
\cite{HocEas88}.

\begin{figure*}
\begin{center}
\resizebox{2.2in}{!}{\includegraphics{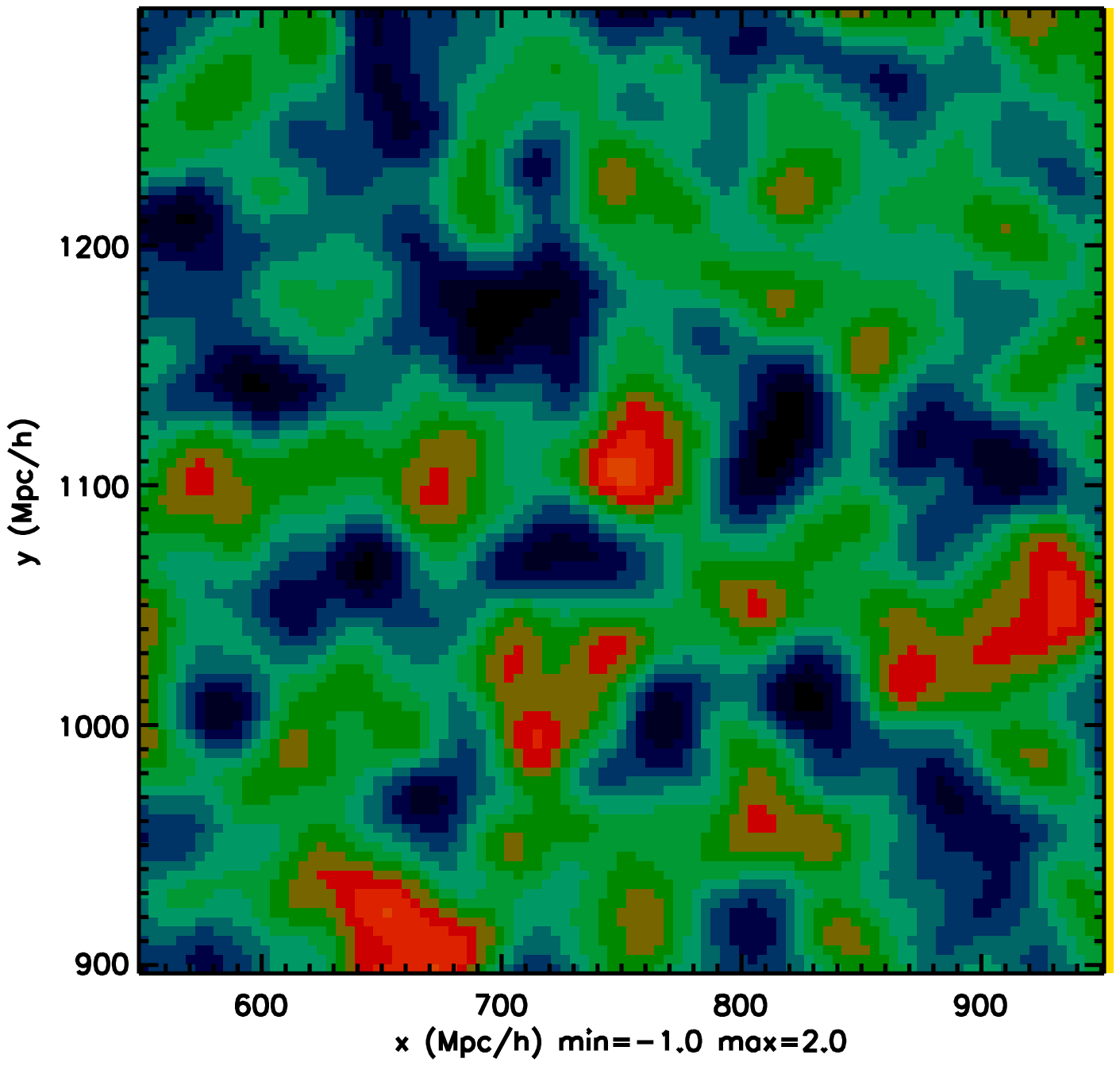}}
\resizebox{2.2in}{!}{\includegraphics{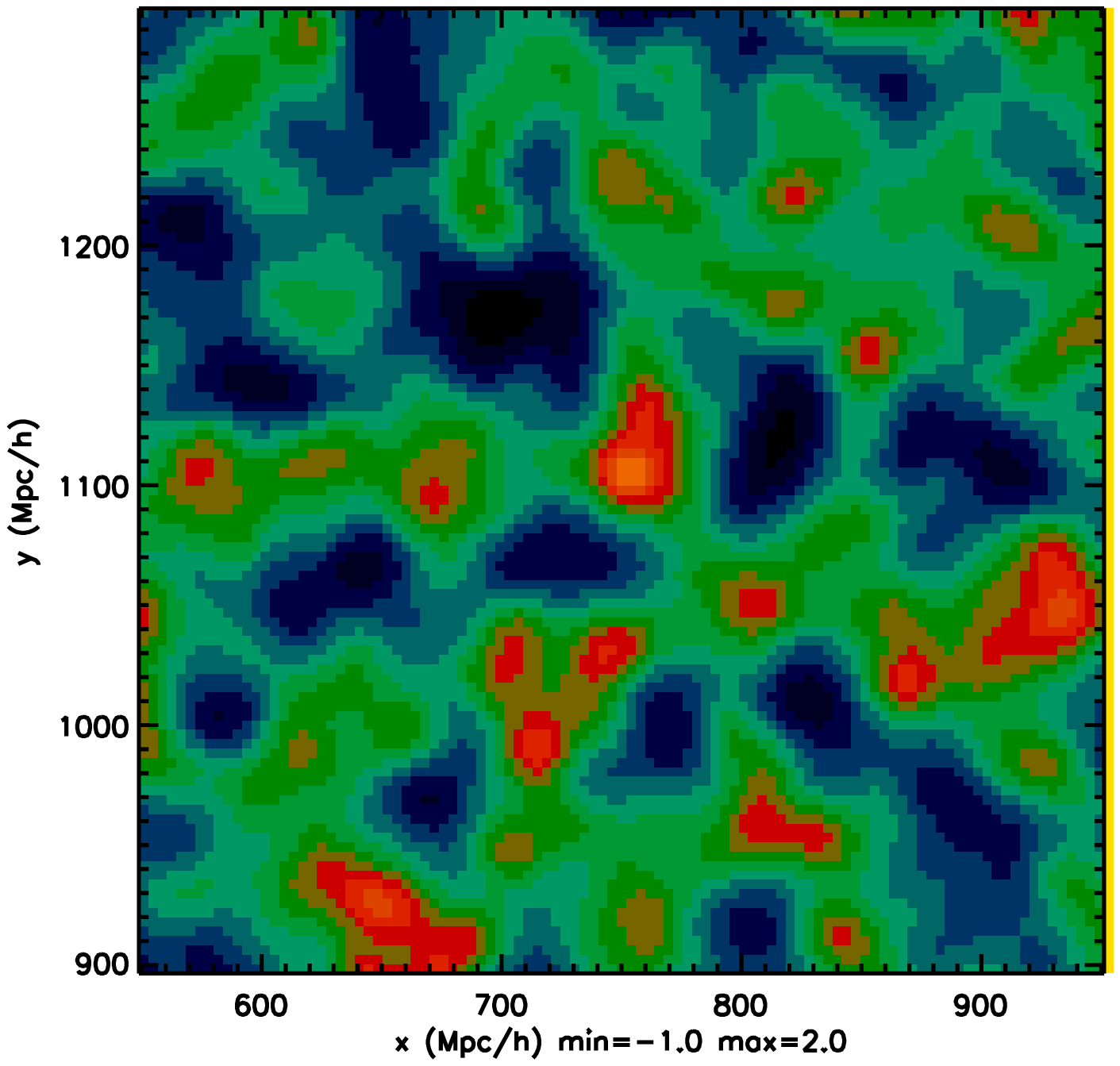}}
\resizebox{2.2in}{!}{\includegraphics{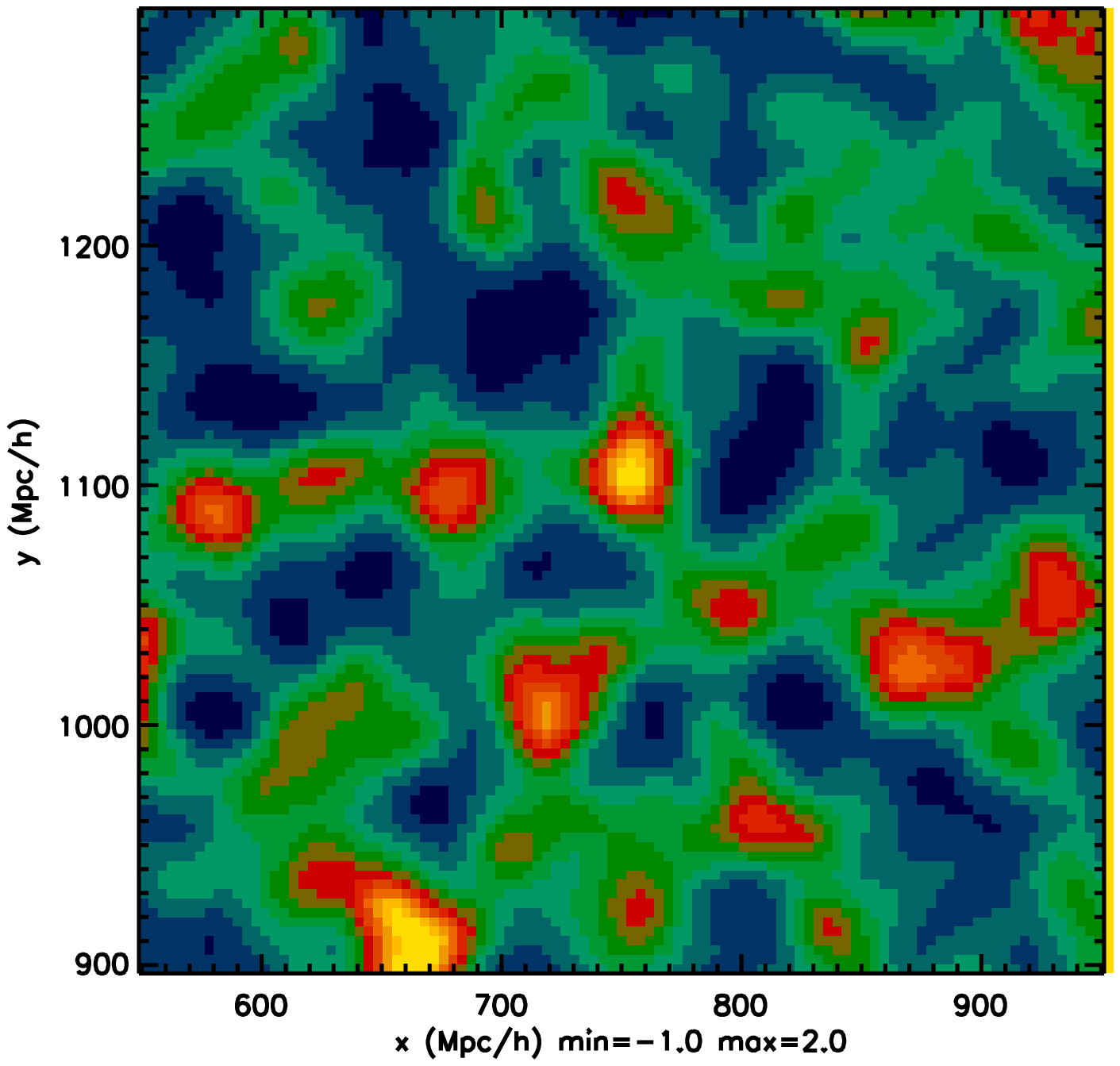}}
\end{center}
\vspace{-0.2in}
\caption{A thin slice through a simulation showing the initial
(left), reconstructed (middle) and final (right) density fields all smoothed
with a Gaussian of $10\,h^{-1}$Mpc.  Each slice is centered on the (final)
position of a halo of mass $4\times 10^{14}\,h^{-1}M_\odot$.
Note the final field has sharper, more pronounced peaks than either the
initial or reconstructed density fields, though the reconstructed field
still has more prominent peaks than the initial field.}
\label{fig:slice}
\end{figure*}

%\begin{figure}
%\begin{center}
%\resizebox{3.5in}{!}{\includegraphics{vec_wfin_4.15e+14_Az0_z0.0_10.0_512_smth_cl.eps}}
%\end{center}
%\vspace{-0.2in}
%\caption{The displacement vector field, $\mathbf{s}(\mathbf{q})$, plotted
%over a thin slice of the (final) density field around the same halo as in
%Fig.~\protect\ref{fig:slice}. {\bf FIX!???!!}}
%\label{fig:vectors}
%\end{figure}

Fig.~\ref{fig:example} shows an example of reconstruction, for the
$\Lambda$CDM model.  By $z=0$ non-linear evolution has partially washed out
the peak in the matter correlation function (solid line).  However applying
reconstruction with $R=5$ or $10\,h^{-1}$Mpc restores much of the original
signal.  Fig.~\ref{fig:slice} shows reconstruction at the level of the
density fields, for a thin slice through a piece of one of our simulations
centered on a halo of mass $4\times 10^{14}\,h^{-1}M_\odot$.
We see that reconstruction has `reversed' the formation of collapsed
structures, and yields a field that is visually similar to the initial
density field.  Note the final field has sharper, more pronounced peaks
than either the initial or reconstructed density fields, though the
reconstructed field still has more prominent peaks than the initial field.

\subsection{Lagrangian perturbation theory}

Reconstruction naturally lends itself to a description in term of Lagrangian
perturbation theory, which we briefly review here.
The Lagrangian description of structure formation \cite{Buc89,Mou91,Hiv95}
relates the current (or Eulerian) position of a mass element, $\mathbf{x}$, to
its initial (or Lagrangian) position, $\mathbf{q}$, through a displacement
vector field $\mathbf{\Psi}(\mathbf{q})$,
\begin{equation}
  \mathbf{x} = \mathbf{q} + \mathbf{\Psi}(\mathbf{q}) \,.
\end{equation}
The displacements can be related to overdensities by \cite{TayHam96}
\begin{equation}
  \delta(\mathbf{k}) = \int d^3q\ e^{-i \mathbf{k}\cdot \mathbf{q}}
  \left(e^{-i \mathbf{k}\cdot \mathbf{\Psi}(\mathbf{q})} - 1\right) \quad .
\label{eqn:lptdensity}
\end{equation}
Analogous to Eulerian perturbation theory, LPT expands the displacement in
powers of the linear density field, $\delta_L$,
\begin{equation}
  \mathbf{\Psi} = \mathbf{\Psi}^{(1)} + \mathbf{\Psi}^{(2)} + \cdots \; ,
\label{eqn:psiexp}
\end{equation}
with $\mathbf{\Psi}^{(n)}$ being $n^{\rm th}$ order in $\delta_L$.
First order in LPT is equivalent to the well-known Zel'dovich approximation.

In the simulations the rms ($1D$) displacement goes from $6.1\,h^{-1}$Mpc
at $z=0$ to $3.9\,h^{-1}$Mpc at $z=1$, in excellent agreement with the
expectations of the Zel'dovich approximation.  In fact the Zel'dovich
rms displacements match those measured in the simulations at the percent
level, better than we would expect given the size of the second order
corrections.

\begin{table}
\begin{tabular}{c|cc|cc|cc|cc}
    & \multicolumn{4}{c|}{$z=0$} & \multicolumn{4}{c}{$z=1$} \\ \hline
    & \multicolumn{2}{c|}{Shifted} & \multicolumn{2}{c|}{Displaced}
    & \multicolumn{2}{c|}{Shifted} & \multicolumn{2}{c}{Displaced} \\
$R$ & \  Sim & 1LPT &\ Sim & 1LPT & \  Sim & 1LPT &\ Sim & 1LPT \\ \hline
 5  &   5.82 & 5.39 & 3.35 & 1.95 &   3.76 & 3.41 & 2.16 & 1.23 \\
10  &   4.92 & 4.80 & 3.40 & 2.80 &   3.20 & 3.04 & 2.19 & 1.77 \\
15  &   4.39 & 4.34 & 3.72 & 3.36 &   2.85 & 2.75 & 2.38 & 2.13 \\
20  &   3.99 & 3.97 & 4.00 & 3.77 &   2.59 & 2.52 & 2.56 & 2.38 \\
25  &   3.67 & 3.67 & 4.24 & 4.07 &   2.39 & 2.32 & 2.71 & 2.58 \\
30  &   3.40 & 3.41 & 4.45 & 4.32 &   2.21 & 2.16 & 2.84 & 2.73
\end{tabular}
\caption{The rms displacements of the ``shifted'' and ``displaced''
particles at $z=0$ and $z=1$ as a function of the smoothing scale $R$. 
First order LPT correctly predicts the observed displacements (at the 10\%
level), with the agreement improving as the smoothing scale increases.}
\label{tab:Sigma}
\end{table}

Using Eq.~(\ref{eqn:lptdensity}) the power spectrum is
\begin{equation}
  P(k) = \int d^3q\ e^{-i \mathbf{k}\cdot \mathbf{q}} 
  \left( \left\langle e^{-i k_{i} \Delta\Psi_{i}(\mathbf{q})}
         \right\rangle-1\right) \,,
\label{eqn:lptpk}
\end{equation}
where $\mathbf{q} = \mathbf{q}_1 - \mathbf{q}_2$, and
$\Delta\mathbf{\Psi}=\mathbf{\Psi}(\mathbf{q}_1)-\mathbf{\Psi}(\mathbf{q}_2)$.
Expanding the exponential in powers of $\mathbf{\Psi}$ and using
Eq.~(\ref{eqn:PsiExpand}) reproduces the results of ``standard'' perturbation
theory.  However, following \cite{Mat08a}, if we use the cumulant expansion
theorem to expand the exponential and expand the resulting powers of
$\mathbf{k}\cdot\Delta\mathbf{\Psi}$ using the binomial theorem we have
two types of terms: those where the $\mathbf{\Psi}$ are all evaluated at the
same point (which we can take to be the origin) and the rest.
Leaving the first set of terms exponentiated
while expanding the second set of terms in powers of $\mathbf{\Psi}$,
we find
\begin{equation}
  P(k) = e^{-k^2\Sigma^2/2}\left\{ P_L(k) + \cdots \right\}
\end{equation}
where $P_L(k)$ is the linear theory power spectrum, $\Sigma$ is proportional
to the rms Zel'dovich displacement
(i.e.~final minus initial particle positions to linear order)
\begin{equation}
  \Sigma^2 = \frac{1}{3\pi^2} \int dq\ P_L(p)
\end{equation}
and explicit expressions for the higher order terms may be found in
\cite{Mat08a} and Appendix~\ref{app:LPT}.
The exponential prefactor describes the broadening of the acoustic peak seen
in Fig.~\ref{fig:example}, some of the additional terms lead to a slight
change in the peak position \cite{CroSco08,PadWhiCoh09,PadWhi09}.
The rms displacement of an individual particle is $\Sigma/\sqrt{2}$.

The effects of the exponential prefactor are most easily seen by considering
the correlation function.
Furthermore, Lagrangian perturbation theory, like several other perturbation
theory schemes, performs better at predicting the large-scale correlation
function than the power spectrum, since it fails to accurately predict
broad-band power which contributes at small $r$
\cite{Mat08a,PadWhi09}.
For these reasons, we shall present most of our comparisons between theory
and simulation in configuration space, i.e.~we shall present
\begin{equation}
  \xi(r) = \int\frac{d^3k}{(2\pi)^3}\ P(k) j_0(kr)
         = \int\frac{dk}{k}\ \Delta^2(k)j_0(kr)
\end{equation}
with $j_0(x)=\sin(x)/x$ the spherical Bessel function of order zero.
This comparison also has the advantage of more clearly emphasizing the
acoustic feature, which can be easily seen as a single peak in $\xi(r)$
at $r\sim 110\,h^{-1}$Mpc.
For presentation purposes we have smoothed all of the correlation functions
by $3\,h^{-1}$Mpc before plotting them -- this reduces high frequency noise
in the N-body simulations but has a minimal impact on the shape of the curves
since this smoothing adds in quadrature to the $\sim 10\,h^{-1}$Mpc intrinsic
width of the features.  Observationally one could achieve similar effects by
using broad but overlapping $r$ bins.

A second interesting statistic is the cross-spectrum
between the linearly evolved initial field, $\delta_L$ and the
fully evolved final field, $\delta_f$,
\begin{equation}
  G_f(k)\equiv
  \frac{\left\langle\delta_L(k)\delta_f^\star(k)\right\rangle}{P_L}
  \quad ,
\label{eqn:Gkdef}
\end{equation}
sometimes referred to as the propagator \cite{CroSco08}.
The relevant physics in this case is more cleanly visualized in Fourier space,
since it shows the decorrelation between the initial field and the processed
field which becomes a convolution in configuration space.
Fits to numerical simulations \cite{ESW07} and a variety of analytic arguments
\cite{Bha96,ESW07,CroSco08,Mat08a,Seo08,PadWhiCoh09,PadWhi09},
including Lagrangian perturbation theory, suggest that
\begin{equation}
  G_f(k) \simeq e^{-(k\Sigma)^2/4} + \cdots
\label{eqn:Sigdef}
\end{equation}
i.e.~that the damping is half as strong as in the power spectrum
(see Appendix \ref{app:LPT}
for expressions beyond leading order).

It is straightforward to repeat these steps for the reconstructed field
\cite{PadWhiCoh09}. We assume that the density field is smoothed on a large
enough scale that $\mathbf{s}$ can be approximated as
$\mathbf{s}=-i(\mathbf{k}/k^2)\delta_L(\mathbf{k}){\cal S}(k)$.
We can then compare the three contributions to the power spectrum 
($P_{ss}$, $P_{dd}$ and $P_{sd}$) to get the reconstructed power spectrum \cite{PadWhiCoh09}
\begin{equation}
\begin{array}{ll}
  P_r(k)
  &= \left\{e^{-k^2\Sigma^2_{ss}/2} {\cal S}^2(k) \right. \\
  & + 2 e^{-k^2 \Sigma^2_{sd}/2}
        {\cal S}(k)\bar{{\cal S}}(k) \\
  & + \left. e^{-k^2\Sigma^2_{dd}/2} \bar{\cal S}^2(k) \right\} P_L(k) + \ldots
\end{array}
\label{eqn:Pr_mass}
\end{equation}
where $\bar{\mathcal{S}}\equiv 1-\mathcal{S}$, and as before, the higher order 
terms are Appendix~\ref{app:LPT}. There are now three smoothing terms,
($\Sigma_{ss}$, $\Sigma_{dd}$ and
$\Sigma_{sd}$) defined by 
\begin{equation}
  \Sigma_{ss}^2 \equiv  \frac{1}{3\pi^2} \int dq\ P_L(p) {\cal S}^{2}(p)\,,
\end{equation}
\begin{equation}
  \Sigma_{dd}^2 \equiv  \frac{1}{3\pi^2} \int dq\ P_L(p) \bar{\cal S}^{2}(p)\,,
\end{equation}
and $\Sigma_{sd}^{2} = (\Sigma_{ss}^2 + \Sigma_{sd}^2)/2$
(see Table \ref{tab:Sigma}).
As pointed out in Ref.~\cite{PadWhiCoh09}, all of these smoothing scales are
smaller than the nonlinear smoothing $\Sigma$, explaining why the acoustic
feature is sharpened after reconstruction.
A related calculation (see Appendix~\ref{app:LPT}) yields the propagators
\begin{eqnarray}
  G_f &=& e^{-k^2\Sigma^2/4} + \ldots \\
  G_d &=& e^{-k^2\Sigma_{dd}^2/4}\ \bar{\mathcal{S}} + \ldots \\
  G_s &=& e^{-k^2\Sigma_{ss}^2/4}\ \left[ -\mathcal{S}\right] + \ldots \\
  G_r &\equiv& G_d-G_s \quad ,
\end{eqnarray}
with the higher order terms in the Appendix.

\begin{figure}
\begin{center}
\resizebox{3.5in}{!}{\includegraphics{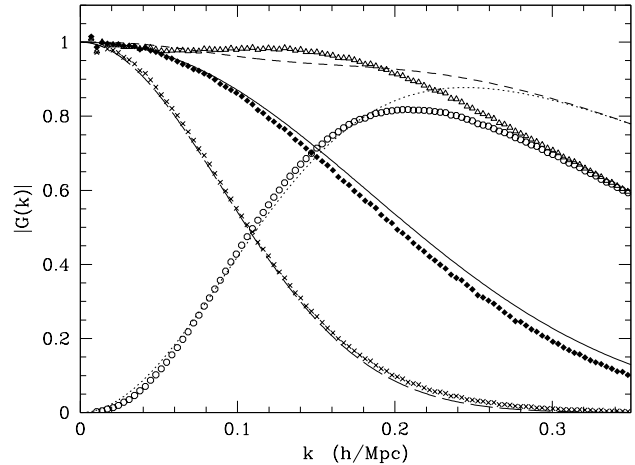}}
\end{center}
\vspace{-0.2in}
\caption{The cross-correlation between the linearly evolved initial field
and the fully evolved final field, displaced field, shifted field and
the reconstructed field for $\Lambda$CDM at $z=0$ (see text).
The points show the results of N-body simulations while the lines show the
predictions from Lagrangian perturbation theory \protect\cite{PadWhiCoh09}.
The solid line and diamonds represent $G_f(k)$, the dotted line and circles
represent $G_d(k)$, long-dashed line and crosses represent $G_s(k)$
and short-dashed line and triangles represent $G_r(k)$.
We have used a smoothing of $R=10\,h^{-1}\,$Mpc.}
\label{fig:Gk_mass}
\end{figure}

\subsection{Comparison with simulations}

\begin{figure}
\begin{center}
\resizebox{3.5in}{!}{\includegraphics{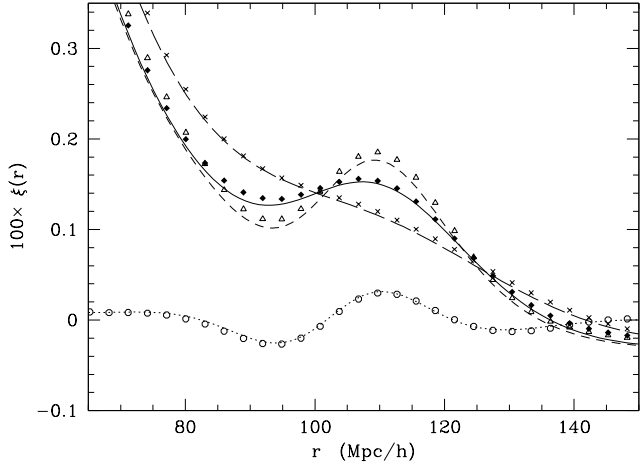}}
\end{center}
\vspace{-0.2in}
\caption{The correlation functions of the fully evolved final, displaced,
shifted, and reconstructed mass fields for $\Lambda$CDM at $z=0$ (see text).
As in Fig.~\protect\ref{fig:Gk_mass} the points show the results of N-body
simulations while the lines show the predictions from Lagrangian perturbation
theory \protect\cite{PadWhiCoh09}.
Solid line and diamonds represent $\xi_f$, dotted line and circles represent
$\xi_d$, long-dashed line and crosses represent $\xi_s$, and
short-dashed line and triangles represent $\xi_r$.
We have used a smoothing of $R=10\,h^{-1}\,$Mpc.}
\label{fig:Xi_mass}
\end{figure}

To begin we compare the predictions of perturbation theory for the propagator
to calculations of the same quantity in N-body simulations.  This isolates the
damping behavior from the mode-coupling \cite{PadWhi09}.
Fig.~\ref{fig:Gk_mass} shows the different contributions to the reconstructed
propagator.
The theoretical predictions for $G_f$ are in reasonably good agreement with the
results, with the theory showing slightly weaker damping than the simulations.
(Small changes to the theoretically predicted $\Sigma$ can bring the results
into much better agreement, but we will not make such {\it ad hoc\/} changes
here.)
The agreement is somewhat worse for some pieces of the reconstructed
propagator.
In particular the simulations show that the reconstructed field retains better
memory of its initial conditions ($G_r\approx 1$) at intermediate scales
than LPT predicts, with perturbation theory giving too much power at high $k$.
The over-prediction at high $k$ is not of particular concern, since at these
scales the dimensionless power exceeds unity and we would expect perturbation
theory to be breaking down.
Out to $k\simeq 0.2\,h\,{\rm Mpc}^{-1}$, where $\Delta^2\sim 1$, perturbation
theory agrees with the simulations at the better than 10\% level!
We emphasize that this level of agreement comes from the inclusion of the
2$^{\rm nd}$ order contributions, with the dominant correction coming from
the $R_1$ term (see Appendix \ref{app:LPT}).

Fig.~\ref{fig:Xi_mass} shows the corresponding figure for the correlation
functions, broken down into the same components.  Note the excellent agreement
for the displaced and shifted fields, but less good agreement for the final
and reconstructed fields.
In this figure the level of agreement between $\xi_f$ and the theory is worse
than the comparable figure in \cite{Mat08b}.  This is most likely due to the
lower redshift and different cosmology we have chosen
(see also \cite{CarWhiPad09}).
The sense of the disagreement in both $\xi_f$ and $\xi_r$ is the same however,
indicating that the Lagrangian perturbation theory of reconstruction is
working better in a differential than absolute sense.
As above, a small change in the relevant $\Sigma$ could slightly improve the
agreement with simulations, which may argue for leaving $\Sigma$ as a free
parameter when fitting to data.  We will not pursue such modifications further
here.

\section{Reconstruction II: Biased tracers}

Unfortunately we don't directly measure the mass field in galaxy surveys,
we measure the distribution of biased tracers.  Here we investigate how
the biasing of the tracers affects reconstruction.
Rather than attempt a `realistic' galaxy model, we shall concentrate on
mass limited samples of halos when comparing LPT to the simulations.
None of the essential aspects are lost with this simplification.

Reconstruction assumes that we can estimate the appropriate shifts from our
smoothed, biased, density field.  This requires that the smoothed halo field
be a multiple of the smoothed mass field with known constant of proportionality
(the bias).  In the simulations we estimate the bias from the $k\approx 0$
limit of the propagator, in observations it would need to be determined in a
different manner.

\begin{figure}
\begin{center}
\resizebox{3.5in}{!}{\includegraphics{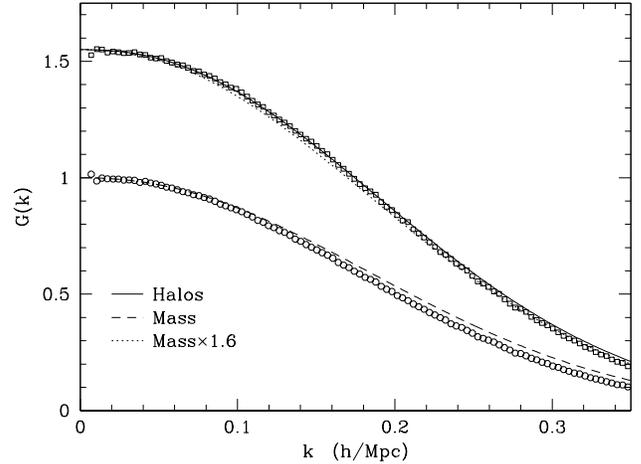}}
\end{center}
\vspace{-0.2in}
\caption{The cross-correlation between the linearly evolved initial field
and the fully evolved final field for the mass (dashed line and circles)
and for halos above $10^{13}\,h^{-1}M_\odot$ (solid line and squares)
in $\Lambda$CDM at $z=0$.
The dotted line shows the mass propagator multiplied by $b\simeq 1.6$.}
\label{fig:Gk_hybrid}
\end{figure}

If we keep the denominator in Eq.~(\ref{eqn:Gkdef}) as the linear mass power
spectrum, the lowest order modification to the propagator is to multiply by
the linear bias of the tracer.
The gross shape of $G(k)$ is unaltered, since the exponential damping is
unchanged, being generated by the velocities which are sourced by the mass
field not the halo field.
At higher order, the cross terms between the linear and $n$th order terms
are modified and introduce an additional dependence on the bias \cite{Mat08b},
as shown in Fig.~\ref{fig:Gk_hybrid}.
LPT predicts that the halo propagator falls slightly more slowly to high
$k$ than the mass propagator and the decline is slower the higher the mass
threshold.
This means that the halo propagator departs more from the Gaussian form than
the mass propagator.
It is possible that this is related to the special locations in the velocity
field that rare, highly biased peaks occupy (e.g.~\cite{BBKS,PerSch08}).
However, the difference is small, as shown explicitly in
Fig.~\ref{fig:Gk_hybrid}.

\begin{figure}
\begin{center}
\resizebox{3.5in}{!}{\includegraphics{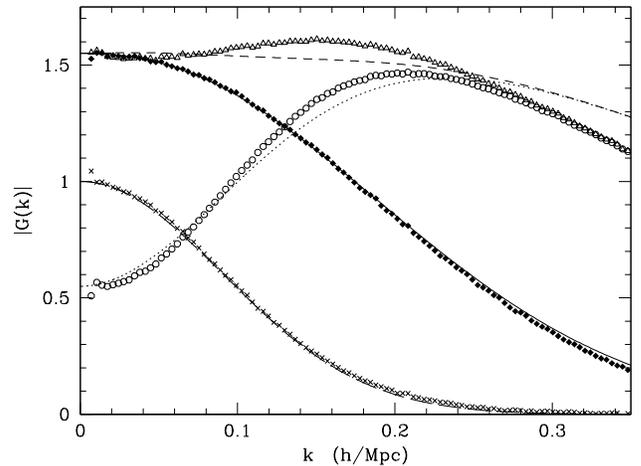}}
\end{center}
\vspace{-0.2in}
\caption{The cross-correlation between the linearly evolved initial field
and the evolved, displaced, shifted and reconstructed fields for halos
above $10^{13}\,h^{-1}M_\odot$.
Diamonds and the solid line show the final field, crosses and the long-dashed
line the shifted field, circles and the dotted line the displaced field and
the triangles and short-dashed line the reconstructed field.}
\label{fig:Gk_halo}
\end{figure}

Fig.~\ref{fig:Gk_halo} shows the different propagators for halos more massive
than $10^{13}\,h^{-1}M_\odot$ in the simulations and in theory.  As was the
case for the mass, the asymptote at high $k$ is not well determined by the
theory but the agreement at low $k$ is quite good.  $G_s$ is the same as for
the mass, and again the agreement between simulation and theory is good.
The match between simulations and theory for $G_f$ is quite good.
Perturbation theory is correctly predicting the low $k$ asymptote of $G_d$,
which is no longer zero but $b-1$, though it doesn't match the shape as well
as for the mass.  Once more the N-body simulations predict a $G_r$ which
increases slightly at intermediate $k$ and is above the theory for
$k\simeq 0.1-0.2\,h\,{\rm Mpc}^{-1}$.

To lowest order (see Appendix \ref{app:LPT} for $2^{\rm nd}$ order
contributions) the reconstructed field has
\begin{equation}
\begin{array}{ll}
  P_r^{(0)}(k)
  &= P_L(k)\left\{e^{-k^2\Sigma^2_{ss}/2} {\cal S}^2(k) \right. \\
  & + 2 e^{-k^2 \Sigma^2_{sd}/2}
    \left[  {\cal S}(k)\bar{{\cal S}}(k) + (b-1){\cal S}(k) \right] \\
  & + \left. e^{-k^2\Sigma^2_{dd}/2}
    \left[ \bar{\cal S}^2(k)+2(b-1)\bar{{\cal S}}(k)+(b-1)^2\right] \right\}
\end{array}
\end{equation}
which reduces to Eq.~(\ref{eqn:Pr_mass}) in the limit $b\to 1$.  Note that
$P_r^{(0)}(k)\to b^2 P_L(k)$ as $k\to 0$, as expected, and
$P_r^{(0)}(k)\to b^2 P_L(k)\exp[-k^2\Sigma^2/2]$ in the limit that
$\Sigma_{ss}=\Sigma_{dd}=\Sigma_{sd}$.

\begin{figure}
\begin{center}
\resizebox{3.5in}{!}{\includegraphics{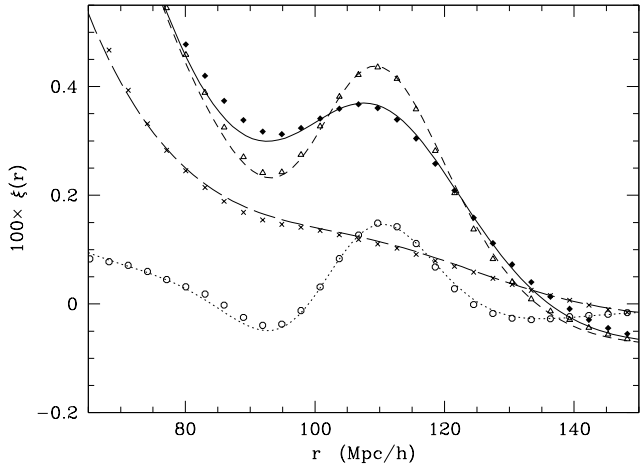}}
\end{center}
\vspace{-0.2in}
\caption{The correlation functions for the evolved, displaced, shifted and
reconstructed fields for halos above $10^{13}\,h^{-1}M_\odot$.
Diamonds and the solid line show the final field, crosses and the long-dashed
line the shifted field, circles and the dotted line the displaced field and
the triangles and short-dashed line the reconstructed field.}
\label{fig:Xi_halo}
\end{figure}

Fig.~\ref{fig:Xi_halo} shows how well this expression, plus the $2^{\rm nd}$
order contributions, matches the simulations.  As with Fig.~\ref{fig:Xi_mass},
the agreement is overall quite good, slightly better than for the mass in
the case of $\xi_f$ and $\xi_r$.  As in that case, a slight increase in the
$\Sigma$ can improve the agreement somewhat, but we have left the theoretical
predictions unchanged.

\begin{figure}
\begin{center}
\resizebox{3.5in}{!}{\includegraphics{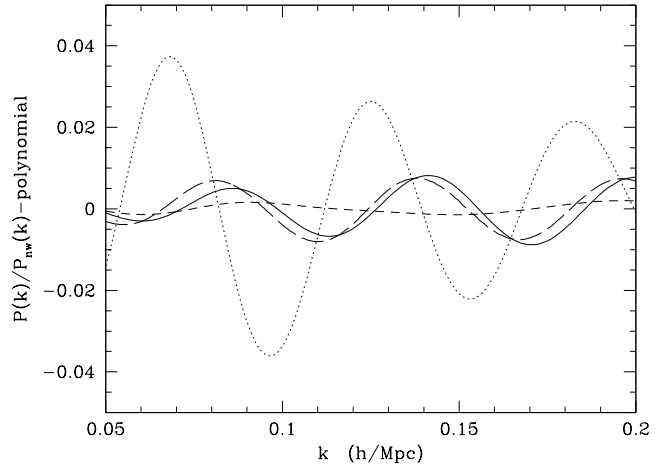}}
\end{center}
\vspace{-0.2in}
\caption{The out-of-phase pieces of the power spectrum of halos more
massive than $10^{13}\,h^{-1}M_\odot$ as predicted by perturbation theory.
To emphasize the oscillations, each spectrum has been divided by the
``no wiggle'' form of Ref.~\protect\cite{EisHu98} and has had a $4^{\rm th}$
order polynomial (in $k$) subtracted.
The dotted line shows the linear theory (divided by 2).  The solid line is
the out-of-phase or mode-coupling pieces of $P_f$, which can be compared to
$dP_L/d\ln k$ (long-dashed line) \protect\cite{PadWhi09}.
The short-dashed line shows that reconstruction reduces the amplitude of
the out-of-phase terms and hence the change in the location of the acoustic
peak in $\xi(r)$.}
\label{fig:Pk_halo}
\end{figure}

Just as with the matter field, the smearing of the acoustic peak is reduced
by reconstruction.  In fact there is relatively little difference between the
biased and unbiased tracers in this respect.

\section{Change in the peak location}

The above sections demonstrate that the LPT provides a good description of
how reconstruction reduces the smoothing of the acoustic feature, both for
the dark matter and halos.
Recent simulations \cite{Seo08} have also found that reconstruction corrects
the $\sim 0.5\%$ change in the acoustic scale caused by nonlinear evolution.
It is therefore interesting to see how this is manifest within Lagrangian
perturbation theory.

In perturbation theory the change in the acoustic peak location comes about
because there are second-order corrections to $P(k)$ which are out-of-phase
with the linear theory oscillations
\cite{ESW07,CroSco08,Mat08a,Seo08,PadWhiCoh09}.
The out-of-phase component is quite similar to the derivative of $P_L$ so,
by Taylor's theorem, this addition is akin to a change in the characteristic
frequency of the oscillation.
We consider the analogous terms for the reconstructed power spectrum below, 
in order to explain how reconstruction suppresses such changes.

These out-of-phase components come about because of the structure of the
mode-coupling terms (the $Q_n$ in the notation of Appendix \ref{app:LPT}
and Ref.~\cite{Mat08b}),
and this structure is modified by reconstruction in such a way as to
reduce the amplitude of the out-of-phase contribution \cite{PadWhiCoh09}.
Figure \ref{fig:Pk_halo} shows the out-of-phase terms, with the broad-band
shape removed to focus on the oscillatory structure, compared to the in-phase
acoustic signature in the linear theory.  Note that the modification of the
mode-coupling terms detailed in the Appendix drastically reduces the amplitude
of the out-of-phase terms in the reconstructed spectrum, and hence the change
in the acoustic scale.
This explains why the change in the peak location seen in simulations is
reduced by reconstruction.

\section{Discussion} \label{sec:discuss}

Acoustic oscillations in the photon-baryon fluid prior to decoupling leave
an imprint both in the cosmic microwave background anisotropy power spectrum
and the matter power spectrum.  A comparison of these features at different
redshifts provides one of the most promising routes to constraining the
expansion history of the Universe.  Unfortunately at low redshift, where
the accelerated expansion of the Universe is strongest, non-linearities
wash out much of the acoustic information.

Recently the authors of Ref.~\cite{ESSS07} proposed a method for recovering
much of the lost information, or reconstructing the acoustic peak.  
Unfortunately, the method is inherently non-linear and therefore difficult
to understand analytically.
A study of this problem in Lagrangian perturbation theory \cite{PadWhiCoh09},
for the mass field, shed some light on how the algorithm resulted in tighter
constraints on the acoustic scale, but the quantitative validity of Lagrangian
perturbation theory is questionable (see e.g.~\cite{CarWhiPad09} for a recent
survey) and we typically study biased tracers of the mass.

We have validated and extended the analytic insights developed in
\cite{PadWhiCoh09}, computing a variety of statistics of both the mass density
field and the dark matter halo density field using Lagrangian perturbation
theory which we then we then compare to the same quantities measured in a
large suite of N-body simulations.

As emphasized in \cite{PadWhiCoh09}, reconstruction does not generate the
initial power spectrum or correlation function, but it does serve to sharpen
the peak and reduce the change in the peak location associated with
non-linearity.
We demonstrate explicitly that both of these points remain true for biased
tracers.  The amount by which the non-linear smearing is reduced is comparable
for biased tracers and for the mass, since it is generated by bulk flows which
are sourced by the mass density independent of the form of the tracer.  The
fact that peaks form in special locations in the density field appears to have
a very small effect.  The reduction in the peak location change due to
reconstruction is at least as dramatic for biased tracers as for the mass,
with the out-of-phase component responsible for the change being reduced in
amplitude by the process of reconstruction.
A discussion of by how much the peak position changes depends on a detailed
description of the fitting methodology and the sample under consideration,
but if we model the observed spectra as in \cite{PadWhi09} we find that
reconstruction reduces the position change by a factor of $2-4$ for
moderately biased tracers like those investigated here.

We conclude that Lagrangian perturbation theory, while not perfect, provides a
good framework for thinking about reconstruction.  It explains in a natural way
how reconstruction works, and how it achieves a reduction in the smearing and
position of the acoustic peak generated by non-linear evolution.
The predictions of LPT agree to within several percent with the results of
N-body simulations on the large scales most relevant to acoustic oscillations,
for both biased and unbiased tracers.
While not shown explicitly in figures here, perturbation theory becomes an
increasingly good description of the simulations at higher redshift, though
the need for reconstruction beyond $z\simeq 1$ is greatly reduced.

\begin{acknowledgments}
We would like to thank Hee-Jong Seo for useful comments on an early draft
of this paper.
The simulations presented in this paper were carried out using computing
resources of the National Energy Research Scientific Computing Center and
the Laboratory Research Computing project at Lawrence Berkeley National
Laboratory.
MW is supported by NASA and the DoE.
This research was additionally supported by the Laboratory Directed Research
and Development program at Lawrence Berkeley National Laboratory, and by
the Director, Office of Science, of the U.S.  Department of Energy under
Contract No. DE-AC02-05CH11231.
\end{acknowledgments}

\appendix

\section{Beyond leading order} \label{app:LPT}

Lagrangian perturbation theory allows us to compute corrections to the lowest
order expressions for $P(k)$ and $G(k)$ listed in the text.  Here we give the
$2^{\rm nd}$ order contributions, following \cite{Mat08a,Mat08b} and
\cite{PadWhi09}.
The notation and procedure is borrowed heavily from these works,
to which we refer the reader for more details.

Recall that the displacement is expanded in powers of the linear density
contrast, $\delta_L$, as \cite{BouColHivJus95}
\begin{eqnarray}
  \mathbf{\Psi}^{(n)}(\mathbf{k}) &=& \frac{i}{n!} \int
  \prod_{i=1}^n \left[\frac{d^3k_i}{(2\pi)^3} \right] \nonumber \\
  &\times& \ (2\pi)^3\delta^{(D)}\left(\sum_i \mathbf{k}_i-\mathbf{k}\right)
  \nonumber \\
  &\times& \mathbf{L}^{(n)}(\mathbf{k}_1,\cdots,\mathbf{k}_n,\mathbf{k})
  \delta_L(\mathbf{k}_1)\cdots\delta_L(\mathbf{k}_n) \; .
  \label{eqn:PsiExpand}
\end{eqnarray}
where the $\mathbf{L}^{(n)}$ have closed form expressions, generated by
recurrence relations.  For example,
\begin{equation}
  \mathbf{L}^{(1)} = \frac{\mathbf{k}}{k^2}
\end{equation}
is the well known Zel'dovich displacement, which is $1^{\rm st}$ order LPT.

The density field for a biased tracer can be defined by the displacement field
$\mathbf{\Psi}(\mathbf{q})$ and a function of the smoothed initial density
field in Lagrangian space, $F[\delta_{L}(\mathbf{q})]$, as
\begin{equation}
\delta_{\rm obj}({\mathbf x}) = \int d^{3}q F[\delta_{L}(\mathbf{q})]
   \delta^{(3)}_{D}(\mathbf{x} - \mathbf{q} - \mathbf{\Psi}) \,\,,
\label{eqn:deltaobjdef}
\end{equation}
where $\mathbf{x}$ and $\mathbf{q}$ are the Eulerian and Lagrangian positions
and $\delta^{(3)}_D$ is the 3D Dirac $\delta$ function.
The power spectrum for such tracers can then be written as \cite{Mat08a,Mat08b}
\begin{eqnarray}
  P(k) = \int d^3q e^{-i \mathbf{k} \mathbf{q}}
  \left[ \int_{-\infty}^\infty
  \frac{d\lambda_1}{2\pi}\frac{d\lambda_2}{2\pi}
  \widetilde{F}(\lambda_1)\widetilde{F}(\lambda_2) \times \right.\nonumber \\
  \left. \left\langle e^{i\left(\lambda_1 \delta_L(\bq_1) +
  \lambda_2 \delta_L(\bq_2)\right) +
  i \mathbf{k}[\mathbf{\Psi}(\bq_1)-\mathbf{\Psi}(\bq_2)]}\right\rangle - 1
  \right] \,\,,
\label{eqn:Pobjdef}
\end{eqnarray}
where $\bq = \bq_1 - \bq_2$ and $\widetilde{F}$ is the Fourier transform of $F$.
The distribution-averaged derivatives of $F(\lambda)$, $\langle F'\rangle$ and
$\langle F''\rangle$, characterize the bias of the sample under consideration.
Expressions for the case of peaks in the initial density field (i.e.~peaks
bias) can be found in \cite{Mat08b}.
For the halos considered in the text ($M\ge 10^{13}\,h^{-1}M_\odot$) we have
$\langle F'\rangle=0.55$ and $\langle F''\rangle=-0.37$, with large-scale
bias $1.55$.

To obtain the propagator we cross-correlate Eq.~(\ref{eqn:deltaobjdef}) with
a field defined by $\exp(i\lambda \delta_{L})$; $\delta^{n}_L$ is then simply
obtained by taking the $n$-th derivative with respect to $\lambda$ and
setting $\lambda$ to zero \cite{PadWhi09}.  This allows us to follow a
procedure similar to that in Eq.~(\ref{eqn:Pobjdef}).

The algebra now follows through as in \cite{Mat08a,Mat08b} using the cumulant
expansion theorem, and collecting all zero-lag correlators to yield, e.g.
\begin{eqnarray}
  \left\langle \delta_L\delta_{\rm obj}\right\rangle
  &\propto& \int d^3q\ e^{-i\mathbf{k}\mathbf{q}} \nonumber \\
  &\times& \left[ B^{10}_{01} + \frac{i}{2}B^{10}_{02}
  + \langle F'\rangle\left( B^{11}_{01}-i\xi\right) \right]
\end{eqnarray}
where we have omitted the exponential damping terms for brevity and defined
\cite{Mat08b}
\begin{equation}
  \begin{array}{l}
  B^{n_1 n_2}_{m_1m_2} \equiv (-1)^{m_1} \\
  \vphantom{\int_0^1}
  \times \langle [\delta_{L}(\bq_1)]^{n_1} [\delta_{L}(\bq_1)]^{n_2}
  [\mathbf{k}\mathbf{\Psi}(\bq_1)]^{m_1}
  [\mathbf{k}\mathbf{\Psi}(\bq_2)]^{m_2} \rangle_c \,\,, \end{array}
\end{equation}
with $\langle \cdots \rangle_c$ denoting the connected moments.

Straightforward algebra then yields
\begin{eqnarray}
  \left\langle \delta_L\delta_{\rm obj}\right\rangle &\propto&
  P_L+\frac{5}{21}R_1+\frac{3}{7}R_2 \nonumber \\
  &+& \langle F'\rangle\left( P_L+\frac{3}{7}\left\{R_1+R_2\right\}
  \right)
\end{eqnarray}
where \cite{Mat08b}
\begin{equation}
  R_n(k) \equiv \frac{k^3}{(2\pi)^2} P_L \int_0^\infty dr\ P_L(kr)
  \widetilde{R}_n(r)
\label{eqn:Rndef}
\end{equation}
and
\begin{eqnarray}
  \widetilde{R}_1 &=& \int_{-1}^{1} d\mu\ \frac{r^2(1-\mu^2)^2}{1+r^2-2r\mu} \\
  \widetilde{R}_2 &=& \int_{-1}^{1}
  d\mu\ \frac{(1-\mu^2)r\mu(1-r\mu)}{1+r^2-2r\mu}
\end{eqnarray}
while for the power spectrum, omitting the damping terms, \cite{Mat08b}
\begin{eqnarray}
  P_{\rm obj} &\propto& \left(1+\langle F'\rangle\right)^2 P_L
  + \frac{9}{98}Q_1 + \frac{3}{7} Q_2 + \frac{1}{2}Q_3 \nonumber \\
 &+& \langle F' \rangle  \left[ \frac{6}{7} Q_5 + 2 Q_7 \right]
  +  \langle F''\rangle  \left[ \frac{3}{7} Q_8 +   Q_9 \right] \nonumber \\
 &+& \langle F' \rangle^2\left[ Q_9+Q_{11} \right]
  + 2\langle F' \rangle\langle F''\rangle Q_{12} \nonumber \\
 &+& \frac{1}{2}\langle F''\rangle^2 Q_{13}
  + \frac{6}{7} \left(1+\langle F'\rangle\right)^2\left[R_1+R_2\right]
  \nonumber \\
 &-& \frac{8}{21} \left(1+\langle F'\rangle\right) R_1
\end{eqnarray}
with
\begin{eqnarray}
  Q_n(k) &\equiv& \frac{k^3}{(2\pi)^2} \int_0^\infty dr\ P_L(kr)
  \int_{-1}^{+1}d\mu \nonumber \\
  && P_L\left[ k\sqrt{1+r^2-2r\mu} \right] \widetilde{Q}_n(r,\mu)
  \label{eqn:Qtdef}
\end{eqnarray}
and expressions for the $\widetilde{Q}_n$ can be found in \cite{Mat08b}.

Extending these results to reconstruction is now relatively straightforward,
assuming that the smoothed density field can be well approximated by the
linear field.  The shifted field has
$\mathbf{\Psi}=\mathbf{\Psi}^{(1)}\mathcal{S}$
and no higher order contributions, while the displaced field can be
obtained from $\mathbf{\Psi}$ with the replacement
$\mathbf{\Psi}^{(1)}\to \mathbf{\Psi}^{(1)}\left[1-\mathcal{S}\right]$
with $\mathbf{\Psi}^{(n\ge 2)}$ unchanged.  This yields
\begin{eqnarray}
  \left\langle \delta_L\delta_s\right\rangle &\propto&
  -P_L\mathcal{S} \\
  \left\langle \delta_L\delta_d\right\rangle &\propto&
  P_L\bar{\mathcal{S}} + \frac{5}{21}R_1 +\frac{3}{7}R^{(d)}_2 \nonumber \\
  &+& \langle F'\rangle\left( P_L+\frac{3}{7}\left\{R_1+R_2\right\}\right)
\end{eqnarray}
where $\bar{\mathcal{S}}\equiv (1-\mathcal{S})$ and $R^{(d)}_2$ is evaluated
using $P_L\bar{\mathcal{S}}$ inside the integral Eq.~(\ref{eqn:Rndef}).

The power spectrum can be evaluated in a similar fashion, with the three
contributions being
\begin{eqnarray}
  P^{ss} &\propto& P_L \mathcal{S}^2 + \frac{1}{2} Q^{(ssss)}_3
\end{eqnarray}
and
\begin{eqnarray}
  P^{sd}+P^{ds} &\propto&
  -2P_{L} \cs \csb + \frac{3}{7}Q_2^{(1s1s)} + Q^{(sdsd)}_3 \nonumber \\
  &-& \cs \left[ \frac{10}{21}R_1 + \frac{6}{7}R_2^{(d)} \right] \nonumber \\
  &+& \fp \left[-2S P_{L} + 2Q^{(1sds)}_7  \right.
  \nonumber \\
  &-& \left. \frac{6}{7} \cs (R_1 + R_2) \right] \nonumber \\
  &+& \fpp Q^{(1s1s)}_9 
\end{eqnarray}
and
\begin{eqnarray}
  P^{dd} &\propto& P_{L} \csb^2
  + \frac{9}{98} Q_1 + \frac{3}{7}Q^{(1d1d)}_2 + \frac{1}{2}Q^{(dddd)}_3
  \nonumber \\
  &+& \csb \left[ \frac{10}{21}R_1 + \frac{6}{7}R_2^{(d)} \right] \nonumber \\
  &+& \fp \left[ 2P_{L} \csb + \frac{6}{7}Q_5^{(1d11)} +  2Q_7^{(1ddd)} +
  \right. \nonumber \\
  & & \left. \frac{10}{21}R_1+\frac{6}{7}R_2^{(d)} +
      \frac{6}{7} \csb (R_1+R_2)  \right]     \nonumber \\
  &+& \fpp \left[ \frac{3}{7}Q_8 + Q_{9}^{(1d1d)} \right]    \nonumber \\
  &+& \fp^2 \left[ P_{L} + \frac{6}{7}(R_1 + R_2) + Q_{9}^{(1d1d)} + Q_{11}^{(11dd)} \right]   \nonumber \\
  &+& 2 \fp \fpp Q_{12}^{(111d)} + \frac{1}{2} \fpp^2 Q_{13} 
\end{eqnarray}
where we have again omitted the damping terms and the superscripts indicate
which $P_L$ are to be replaced with $P_L\mathcal{S}$, $P_L\bar{\mathcal{S}}$
etc.  For the $Q_n$ there are 4 possible smoothing terms, and we have indicated
no smoothing with a $1$, $\mathcal{S}$ with an $s$ and $\bar{\mathcal{S}}$ with
a $d$.  The first two terms have argument $kr$ and the second have argument
$k\sqrt{1+r^2-2r\mu}$ in Eq.~(\ref{eqn:Qtdef}).  Thus for example
\begin{eqnarray}
  Q_7^{(1sds)}(k) &=& \frac{k^3}{(2\pi)^2} \int_0^\infty dr\ P_L(kr)
  \mathcal{S}(kr) \nonumber\\
  &\times& \int_{-1}^{+1}d\mu\ P_L(ky) \mathcal{S}(ky)\bar{\mathcal{S}}(ky)
  \nonumber \\
  &\times& \widetilde{Q}_7(r,\mu)
\end{eqnarray}
with $y=\sqrt{1+r^2-2r\mu}$.

\onecolumngrid

\bibliography{ms}
\bibliographystyle{apsrev}

\end{document}